# E-Health Management Services in Supporting Empowerment

Muhammad Anshari[1,2]
[1]Universiti Brunei Darussalam School of Business & Economics, anshari.ali@ubd.edu.bn
[2]Universitas Islam Negeri Sunan Kalijaga Yogyakarta, Indonesia

**Abstract**

The Web technology provides healthcare providers the ability to broaden their services beyond the usual practices, and thus provides a particular advantageous environment to achieve complex *e-health* goals. This paper discusses how a Web 2.0 application will help healthcare provider to extend and enhance their services by involving and empowering their customers. Web 2.0 refers to the next generations of Web technology that empowers users to generate contents over Internet. The Web 2.0 also refers to the Web as a platform to perform any task online. The Web 2.0 allows customers to have greater control of information flow from interactions between healthcare providers with its customers and among customers themselves. The study employed quantitative methods to depict expectations of customers in Indonesia towards healthcare services that can offer empowerment and social media and sharing. The questionnaires were fairly distributed to the groups of healthcare staffs and customers (patients) who regularly visit healthcare centres. The survey revealed that features of Web 2.0 in e-health services such as consultation online, sharing in social networks, empowerment in detailing personal health records are highly appreciated by customers. Regardless of the limitations of the survey, the public has responded with a great support for the capabilities of Web 2.0 listed from the questionnaires. The findings provide initial ideas and recommendation for a future direction of prototype of social networks in e-health services.

# Introduction

With the recent advancement of Web 2.0 applications as well as their worldwide adoption, the business world has quickly embraced Web 2.0 in their business processes to take advantage of its network-centric paradigm and interactivity provided to move closer to customers, offering better services as well as strengthening relationship with customers. Web 2.0 became popular following the O'Reilly Web 2.0 conference in 2005. The Web as a platform is the underlying concept of Web 2.0 where customers collectively contribute in providing content and applications (O'Reilley, 2007). It is about social networks to build contextual relationships and facilitate knowledge sharing among customers, and between customers and their service providers (Anshari and Almunawar, 2011). From the customer's perspective, interactive conversation among themselves in order to confirm and even criticize services, products, or performance of firms that they have had experience is very interesting to them. Web 2.0 has driven changes of customers' behaviour in term of engagement, empowerment and intensive interactivity.

In a healthcare environment, the adoption of Web 2.0 brings a promise for both for healthcare providers as well as customers (patients) (Almunawar & Anshari, 2014; Anshari et al., 2013a) It supports better partnerships between patients with their healthcare providers for mutual benefits. Patients' engagement enables healthcare providers to simplify communication and collaboration among patients in creating mutual values by bringing the collective knowledge to achieve patients' satisfaction (Low & Anshari, 2013). On the other hand, patient empowerment has been energized so that they have better control of information flows within social networks they belong to which can affect to improve their health literacy (Almunawar & Anshari, 2012; Anshari et al., 2013b; Anshari et al., 2019a). Web 2.0 can be set as a platform for them to have conversations and interactions among themselves (Mulyani et al., 2019a; Razzaq et al., 2018). The interactions that happen between patients can be considered as a form of empowerment where they can share, discuss and even generate records (contents) of their health related activities (Anshari & Almunawar, 2016a; Hasmawati et al., 2020).

Furthermore, the utilization of information and communications technology (ICT) especially the Internet in the healthcare sector is frequently referred to electronic health or e-health. The main purpose of e-health is to improve healthcare management for mutual benefit between patients and healthcare providers (Anshari & Almunawar, 2016b). One important aspect of e-health is how to manage relationships between a healthcare provider and its patients in order to create greater mutual understanding, trust, and patient involvement in decision-making. This is parallel with the recent Web 2.0 that facilitates customers to generate contents to accommodate both patient-health provider and patient-patient interactions (Anshari et al., 2012a; Anshari & Almunawar, 2015c). Besides, it can be seen as technology and strategy at the same time, bringing promises for the advancement of e-health initiatives. Web 2.0 has brought a possibility to extend the service of e-health service in accommodating social aspect such as enabling patients or patients' families, and the community at large to participate more actively in the process of health promotion and education through a social networking process (Anshari & Almunawar, 2018; Mulyani et al., 2019b).

The main goal of this paper is to introduce a future research direction, which may shape the future of e-health systems. In this paper, we examine customers' expectation concerning the process of initiating Web 2.0 in e-health services so that they can be more proficient in dealing with their own healthcare issues. A survey has been conducted in Malang, Indonesia. The survey focuses on social networks, empowerment, and online health educator as supporting system in an e-health initiative. The output of the survey will be used as a user requirements phase to develop prototypes of Web 2.0 in e-health systems. The rest of this paper is organized as follows: the next section presents the background of the study, followed by the research methodology. We then discuss the results of the study, and set the future direction our research.

## Background of Study

Web 2.0 consists of social and technical aspects (Anshari & Alas, 2015). Social aspects of Web 2.0 are more important than the underlying technologies. Social aspects can transform healthcare organizations from business to consumer activities, engaging more customer-to-customer activities, which in turn can derive massive values from customers' involvement. The main advantages of Web 2.0 are the linkage among people, ideas, processes, systems, contents and other organizational activities (Askool and Nakata, 2010; Anshari, Alas, & Sulaiman, 2019). Web 2.0 may affect healthcare business processes as well, especially those relating to interactions between patients since it offers a new way in engaging, managing and maintaining relationships (Almunawar & Anshari, 2011b).

Web 2.0 tools have brought a possibility to extend the service of healthcare organizations by enabling patients, patients' families and the community at large to participate more actively in the process of health promotion and education through the social networking process. Embedding Web 2.0 in healthcare is a challenging task in order to provide a new meaning in building relationships between patients and healthcare organizations within a social network platform (Almunawar et al., 2012). Furthermore, it is important to encourage patients to share their experiences and related information to achieve better health outcomes.

The use of Web 2.0 in healthcare organization is equivalent to bringing patients' expectation aligned with fashion of ICT in actual healthcare services (Anshari and Almunawar, 2012). The role of Web 2.0 in healthcare organization is now gradually being recognized to the concept known as Health 2.0 or Doctor 2.0. The term became popular following the O'Reilly

Media Web 2.0 conference in 2004. Despite scepticism, the online community using Web 2.0 tools for health continues to grow, and those terms have entered a popular nomenclature (Hughes et al, 20 08). This section discusses interrelated concepts that relevant to the study.

*Social Networks*

The Web 2.0 is an important tool for the development of social network. Social networking can generate a way to strengthen the relationship between organizations and their customers. It can be used as enablers in creating close and long-term relationships between an organization with its customers (Askool, and Nakata, 2010). In addition, Web 2.0 which plays a significant part in the customer relationship management (CRM) transition in organizational, stimulates fundamental changes in consumer behavior (Greenberg, 2009). This revolution is having a broad and deep impact on an interpersonal relationship in all areas, and in healthcare is no exception (Almunawar et al, 2015). The booming number of social networking groups and supports groups for patients on the internet and their influence on health behavior is only beginning to be explored (Rimer and others, 2005) and remains an important area for future research. The concept of a social network defines organization as a system that contains objects such as people, groups, and other organizations linked together by a range of relationships (Askool and Nakata, 2010). Some organizations are building online social networks to engage customers and export ideas, innovations of new services or products, quick feedback, and technologies from people outside the organization (Lafley and Charan, 2008; Anshari et al., 2019b).

In regard to manage the customer relationship, the Internet has become a crucial medium in supporting CRM efforts. Indeed, the Web technology is a powerful channel available for organizations to develop, enhance interactions, and implement relationship practices with customers (Batista, 2010; Ahad et al., 2017). The Web 2.0 is becoming a trend in Web technology and Web design. We are witnessing the acceptance of a second generation of web based communities such as wikis, blogs, and social networking sites which aim to facilitate creativity, collaboration, sharing among users rather than just for email and retrieve some information. Users can own the data on the Web 2.0 site and exercise control over that data (Hinchcliffe, 2006).

Web 2.0, which play a significant part in the CRM transition drives social change that affects all institutions including business and healthcare organizations (Anshari & Almunawar, 2015a). It is a revolution on how people communicate. It facilitates peer-to-peer collaboration and easy access to real time communication. Because much of the communication transition is organized around web-based technologies, it is called Web 2.0 (Greenberg, 2009). With 2.0 technology patients can easily participate in social networks to exchange information and knowledge (Almunawar & Anshari, 2011a). They can share information and knowledge about their diagnoses, medications, healthcare experiences, and other related information. It is often in the form of unstructured communication, which can provide new insights for people involved in the management of health status and chronic care conditions.

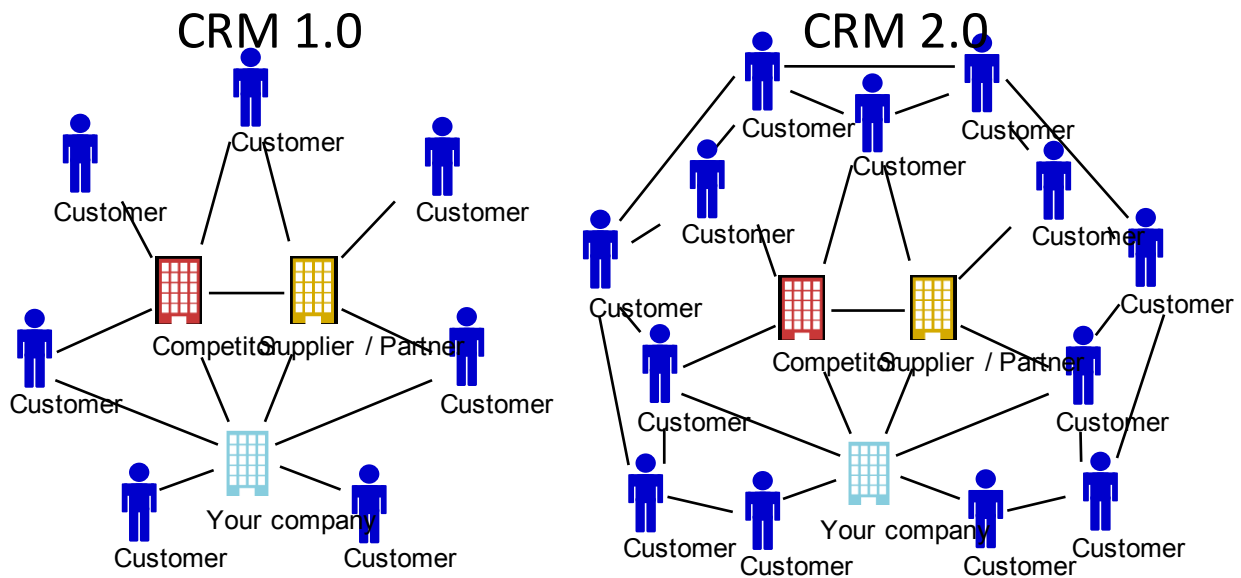

**Figure 1: Evolution of CRM landscape**   Source: Cipriani, 2008

Greenberg (2009) defined Social CRM as a philosophy and a business strategy, supported by a technology platform, business rules, processes, and social characteristics, designed to engage the customer in a collaborative conversation in order to provide mutually beneficial value in a trusted and transparent business environment. It is the company's response to the customer's ownership of the conversation. In this study, the term of Social CRM and CRM 2.0 is used interchangeably. Both share new special capabilities of social media and social networks that provide powerful new approaches to surpass traditional CRM (Anshari & Almunawar, 2015b).

Cipriani described the fundamental changes that Social CRM is introducing to the current, traditional CRM in term of landscape. Figure 1 is the reflection of the evolving CRM 2.0 which is different from CRM 1.0. It is a revolution in how people communicate, customers establish conversation not only with the service provider but it is also with others. Table 1 summarizes the difference of CRM 2.0 from CRM 1.0 based on type of relationship, connection, and how value generated. Relationship type in CRM 1.0 focuses on the individual relationship; Customer to Customer or Customer to Business whereas in CRM 2.0 offers the collaborative relationship and engage a more complex relationship network. Connection type in CRM 1.0 is a limited view of the customer which affect to less informed customer, on the other hand, CRM 2.0 enable for multiple connections allow better understanding and more knowledgeable customer. CRM 1.0 of value creation is constructed from targeted messages, and CRM 2.0 offers diverse value creation even from informal conversation with customers within social networks.

Table 1; Comparison CRM 1.0 and CRM 2.0

| Type | CRM 1.0 | CRM 2.0 |
|---|---|---|
| **Relationship** | Focus on individual relationship (C2C, C2B) | Focus on collaborative relationship (engaging a more complex relationship network) |
| **Connection** | Limited view of the customer & his community preferences, habits, etc | Multiple connections allow better understanding of the customer and his community |
| **Generated Value** | Targeted messages generate value | Conversation generates value |

Source; Cipriani, 2008

Pombriant from Beagle Research Group (2009) analyzed that customers have more control of their relationships with organization a lot more than they were in a few years ago. It is because customers have access to new levels of education, wealth and information. The change is a social change that impacts all organizations.

*E-Government and E-Participation*

E-government procedures provide to citizens easy to handle and fast services, which stepwise replace traditional bureaucratic paper, based contacts between governmental institutions and citizens (Anshari & Lim, 2017). The underlying principle of e-government, supported by an effective e-governance institutional framework, is to improve the internal workings of the public sector by reducing operational costs and transaction times to better integrate workflows and processes and enable effective resource utilization across the various public sector agencies aiming for sustainable solutions through newer technologies. It seeks to establish 'better processes and systems' aimed at more efficiency, effectiveness, inclusion and sustainability (UN, 2012). According to United Nations e-government survey 2012 indicates that global infrastructure access has improved, with the global average ICT index value reflecting an increase in mobile penetration – the global average number of mobile subscriptions per 100 inhabitants is now 88.5. Mobile based technologies have become the most rapidly adapted technologies to provide e-services, playing a pivotal role, especially in developing countries (Anshari, Alas, & Guan, 2017).

According to the Survey, 25 countries have developed separate m-government websites, and 24 countries provide the option of making payments via mobile phones. Many developing countries have adopted citizen inclusion as key in providing "customer"-oriented services. While the Republic of Korea and the Netherlands are the world leaders, Singapore and Kazakhstan are close behind. Europe has the largest share of the top e-participation countries. Leveraging social media for the benefit of e-service uptake is another area where a greater effort can make a difference since currently only 40 % of Member States are using a social networking site (UN, 2012). Table 2 is an e-participation of member states based on survey 2012. The feature shows that Web 2.0 has been embraced in e-participation where there are 18 countries, which citizens can tag, assess, and rank content on the website. Though the number is considered a minor, it directs more participation of people for better services. E-participation is closely related to e-government (Anshari, Alas, & Guan, 2016). E-participation is the generally accepted term referring to ICT-supported participation in processes involved in government and governance. Processes may concern administration, service delivery, decision-making and policymaking. The need for the term has emerged as citizen benefits and values have often received less attention in e-government development than those of the service providers, and the need to distinguish the roles of citizen and customer has become clearer.

Table 2; E-participation of Member States of Survey 2012

| Feature | Number of Countries | % |
|---|---|---|
| Can citizens tag, assess and rank content on the website | 18 | 9% |
| Sections for vulnerable groups such as the poor, illiterate, blind, old, young, immigrants, women, etc | 56 | 29% |
| Provide e-services through or in partnership with third parties such as civil society or the private sector | 71 | 37% |
| Gateway to local or regional government | 97 | 50% |

Source; UNPAN, 2012

E-government and e-participation innovation can be a catalyst in boosting productivity including in the healthcare sector, thereby speeding up the benefit of newer technologies to the people. E-health as part of the e - government program has the potential to secure and raise availability, accessibility and quality of health services. E-health is an essential contributor to improving patient safety (Epsos, 2009). In the last few years many countries have employed ICT in the area of e-health with expansion of related services making it more in tune with people's needs and driving the whole process based on their participation as explained above. In summary, the global trends of e-government has improved in term of infrastructure access, mobile penetration, customer oriented services, social media for the benefit of e-service, and e-participation of people, and those trends are pointing for the future of e-health direction by adopting newest technology or e-participation of the people.

*Empowerment*

Many researchers have discussed the issue of empowerment in healthcare organizations. For instance, empowerment can be analyzed from the perspective of patient-healthcare provider interactions (Skelton, 1997; Paterson, 2001; Dijkstra et al., 2002; van Dam et al., 2003), or from the point of view of the patient alone (Anderson et al., 1995; McCann et al., 1996; Davison et al., 1997; Desbiens et al., 1998; Anshari et al., 2013c), our analysis can encompass both above mentioned perspectives (McWilliam et al., 1997; Golant et al., 2003; Maliski et al., 2004; Anshari et al., 2015). However, research that specifically discusses the issue of empowerment Web 2.0 in the domain of e-health is still very limited. Australia, a pioneer in this matter, has adopted a Personally Controlled Electronic Health Record (PCEHR) system which stands as an example for empowerment through e-health services (NEHTA, 2012). A significant element of patient empowerment has been achieved by allowing them to view their medical information electronically. However, PCEHR has only enabled patients to view their Electronic Medical Record (EMR); it has not utilized other features of Web 2.0, which allow collaboration and conversation between patients or between patients and their healthcare providers. Therefore, the aim of the study is to gather user requirement that facilitates the features of Web 2.0 in order to meet significant challenges in patient empowerment in the domain of healthcare.

The idea of empowerment of patients in healthcare emerged as a response to the rising concern that patients should be able to play a critical role in improving their health. In traditional healthcare practices, a patient is the recipient of care as well as medical decisions. However, a paradigm shift has taken place in healthcare that changes from patients who receive care to those who actively participate in their healthcare. Empowerment is well supported in the healthcare literature, and related to customers and healthcare services over the past decade (Skelton, 1997; Paterson, 2001; Dijkstra et al., 2002; van Dam et al., 2003; Anshari, 2019). In the organization, empowerment implies the provision of necessary tools to staff in order to be able to resolve, on the spot, most problems or questions faced by customers. Besides, staff can deal with customers directly and so, reduce the number of dissatisfied customers who would otherwise have complained, but now, simply switch brands (Low, 2002; Anshari & Almunawar, 2019).

In the traditional paradigm healthcare services, healthcare providers mostly do decision making on treatments. Indeed, the lack of participation of patients in healthcare processes was the main obstacle to the empowerment of patients as customers. Nevertheless, there will always be circumstances in which patients choose to hand over responsibility for decisions about their healthcare to providers due to the difficulty in selecting available options or the time needed to understand the health problem and the options. However, this does not undermine the proposition that customers' empowerment will promote efficiency and that decisions should be made from the perspective of customers (Segal, 1998; Anshari & Sumardi, 2020).

According to McWilliam et al. (1997) empowerment is a result of both interactive and personal processes, where the emergence of 'power' (or potential) is facilitated by caring relationships. Empowerment as an interactive process suggests that power is ''transferred'' by one person to another, whereas empowerment as a personal process suggests that power is

''created'' by and within the person. Although the expected outcome is the same, i.e. the gain of more power over one's life, the nature of the two processes is very different (Aujoulat et al, 2006; Anshari et al, 2021). The first course entails that power can emerge through active co-creation and collaboration in an empowering relationship. In the second case, when the process of empowerment is perceived from the point of view of the customers, it is considered as a process of personal transformation. Next section portrays recent healthcare systems and a penetration of the Internet in Indonesia to give a better idea of the objects research.

*Healthcare System and Internet in Indonesia*

Understanding the object is pivotal to draw the background of study. The study conducted in Malang Indonesia where the city is second largest in East Java Province. It is an urban area that competition intense among healthcare providers, especially among private health providers. In general, National Health Survey Indonesia indicates that the number of community public health centres (Puskesmas) over Indonesia reaches 7,277 units including Puskesmas at a regional level, sub-district level and village level. Moreover, the survey identified that over 30 % of the Indonesian population use Puskesmas to get basic medical services (MoH, 2005). The general decentralization process implemented in 2001 has had many impacts on the health system, even though it was not designed specifically with the health sector in mind. In particular, health financing, health information system, human resources for health and service provision have been affected. Under decentralization, responsibility for healthcare provision is largely in the hands of district governments. Malang is one of the autonomous regions and is the second city in East Java after Surabaya. It has about 73 healthcare centres in 2007 with 800 thousand inhabitants (Sukarelawati, 2011). The private sector is increasingly important in the provision of healthcare in Indonesia, especially in big cities, with wide variations in quality of care.

Initiating Web 2.0 cannot be separated with Internet accessibility and reliability. It is apparently important to figure out the recent Internet adoption in Indonesia and the acceptance of social networking sites by Indonesians. The proliferation of digital media in Indonesia is a function of Internet usage, along with many other local conditions which could encourage or dampen the use of Web 2.0 applications. These include its demographics, Internet-related statistics and its current state of traditional and mobile media, upon which Indonesia is still largely reliant on (SMU, 2011). At year 2000, the Internet users in Indonesia were estimated 4 million (ITU, 2001). However, the Internet users have increased to 30 million or 12 % of total population 242 million in 2010 (ITU, 2010). The number of Internet users in 2011 quickly reached 55 million people; 29 million of them are mobile Internet users (Waizly, 2011). The number of people who use mobile devices has reached 29 million people in 2011, which means that more than 50% Internet users in Indonesia use mobile devices to browse the Internet. Compared to the population of the country which is about 240 million people, this means Indonesia has reached 23% Internet penetration rate and it is dominated by big cities, only 4.1% in rural areas. The Internet penetration is expected to rise significantly in the years to come as technology becomes more affordable. Social media is normally used for sharing and bonding (Thia, 2011). Indonesia is the 4th-largest Twitter nation worldwide. As of January 2011, there were more than 4 million Twitter accounts established by Indonesia users. Almost 20.8 % of online users in Indonesia visited Twitter.com in June 2010 (Comscore, 2010). Currently, there are 44 million Facebook users in the Indonesia, which makes it number 4 in the ranking of all Facebook statistics by Country. Facebook penetration in Indonesia is 18.21% in relation to the country's population and 147.45% in relation to the number of Internet users (Socialbakers, 2012). By looking into the data above, Indonesian are very resilient and large adopters of

Internet technology. It brings a promise for further innovation in term of Internet services including e-health services.

## Methodology

Our main objective is to use Web 2.0 as both a tool and a strategy to empower patients and to link them in social networks in healthcare scenarios. For this purpose we single out potential features offered by Web 2.0 to support empowerment and social networks then incorporate these features in developing out instrument (questionnaire) for our survey. We use the purposive sampling methods in which respondents were intentionally selected from patients, patients' family, and medical staff from hospitals in Malang- Indonesia. Before questionnaires were distributed, we conducted a pilot test with a group of potential users (customers/patients) to make sure it can be easily understood and filled. There were 108 respondents participating in the survey which was conducted in October 2012. Data gathered from the survey was examined, interpreted and eventually converted as requirements to develop a prototype of Web 2.0 in e-health setting.

## Analysis

To analyze the reliability of questionnaire items used in this study, Cronbach's alpha to measure internal consistency. Here N is equal to the number of items, c-bar is the average inter-item covariance among the items and v-bar equals the average variance as in (1). If Cronbach's alpha is greater than or equal to 0.6 then a variable is reliable. Cronbach's Alpha has founded 0.624 from the 19 number of items. It is indicated that they have relatively consistent since it is 0.624 > 0.60, and the survey result is reliable. Furthermore, Table 3 shows the demographic characteristics of the samples. More than 50 % of respondents were from nurse and others. They have completed high school within the ranges of 20 to 40 years old. The level of ICT literacy is between medium to advance user, and 75 % respondents use Internet in daily basis which indicates that the Internet access is not a problem for them.

$$\alpha = \frac{N.\bar{c}}{\bar{v} + (N-1)\bar{c}} \quad (1)$$

Observing the relationship between the variables of rows and columns, we employed chi-square analysis, the Pearson Chi-Square rows and columns Asymp. Sig. (2-sided) that demonstrate the value of probability. From the analysis we found that there is a relationship between educational level and ICT literacy, educational level and access online health services, educational level and payment online services, gender and access online health services, gender and payment of online services.

TABLE 3: DEMOGRAPHIC CHARACTERISTICS OF SAMPLE

| Variable | Component | Percent |
|---|---|---|
| **Employment** | Administrative Health staff | 7.4 % |
| | Doctors | 9.3 % |
| | Nurses | 29.6 % |
| | Others | 53.7 % |
| **Gender** | Male | 27 % |
| | Female | 73 % |
| **Age** | 20 years or younger | 7.4 % |
| | 21 - 30 | 76 % |
| | 31 - 40 | 11 % |
| | 41 - 50 | 3.7 % |
| | 51 years or older | 1.9 % |

| Education | Completed high school | 24.3 % |
|---|---|---|
| | Completed Diploma | 42.12 % |
| | Completed Degree | 31.8 % |
| | Completed Postgraduate | 1.9 % |
| ICT Literacy | Not at all | 7.4% |
| | Basic User | 17.6% |
| | Medium User | 48.1% |
| | Advance User | 26.9% |
| Internet Usage | At least daily | 75% |
| | Weekly | 20.4% |
| | Monthly | 1.9% |
| | Never | 2.8% |

Source: Survey, 2012

Furthermore, there are many issues that we asked to the respondents about health information on the Internet, empowerment features in health service, availability of online health educator, social networks in e-health services, and the effect of those services to improve health literacy and customer satisfaction (Table 4). When we asked to the respondents, do they also use the Internet to obtain health-related information, health science, or health reference? There were 96 % of them used the Internet to get health related information while 4 % were not using it. We followed the question what they were looking for on the Internet, 57% used to look for information about diseases and its treatment, 22% was information about healthy lifestyle, 7% was looking for information about health support and recommendation, and 13% was information about healthcare services.

TABLE 4: SURVEY RESULTS

| Survey's Component | Result |
|---|---|
| **Empowerment** | |
| • View EMR | 69 % |
| • Record health activities online | 75 % |
| **Social networks** | |
| • Discuss health service in social networks | 72 % |
| • Supporting group in social networks | 93% |
| • Discuss with patients a similar condition | 80% |
| **Online Health Educator** | |
| • Consultation online | 83% |
| • Stand by online health educators | 92% |
| **Extended E-Health Services** | |
| • Paying service online | 39% |
| • Emotional & spiritual affect physical | 100% |
| • In overall, improve health literacy | 64% |
| • In overall, improve customer satisfaction | 87% |

Source: Survey, 2012

*Accessibility of Online Health Information*

The concerns of people participation, accessibility or customer centric, and empowerment gain much attention in e-government implementation around the world as discussed in the previous section. Therefore, in this section, the questionnaires concentrate on customers' online accessibility of health information. The first question asked their agreement of the ability of a service that enable them to access their medical records online so that they can monitor their own medical record anywhere and anytime, the study shows that 22%

(strongly agree) and 47% (agree) of the respondents prefer to view and have control of their medical records online.

This is an interesting result, which indicates that they can self-monitor their medical record. The knowledge of the medical status of historical data available in the medical records may lead to improve healthcare awareness and self-managed healthcare. In addition, the online medical record may help them to check and make sure that they have the right health details to avoid miss communications. Further, the online medical records can be used as a guideline in making any decision relating to their own health. Although most of the respondents prefer online medical records, some respondents (26% disagree and 4% strongly disagree) with the idea of online medical records accessibility. This is mainly due to there are tendency of the hesitation of medical staffs to disclose of medical records. The issue becomes challenging task for future research in e-health direction especially the issue of change management.

Further we asked the preference of customers to have ability in recording their own health related activities and habits. There are 75% of respondents like the service to record their daily habits that may affect to their health directly or indirectly. Most respondents were happy if they can record their health-related activities online these activities may include a personal health diary where the respondents can access it, anytime and anywhere, facilitating their daily plans and programs for a healthy lifestyle. This service can be used to monitor their health status and may help in making health decisions later. For instance, customers can record their personal habit like eating, exercises, hobby, weight, even blood pressure. Those records can be used when they make consultation with the medical practitioners. Though there are 25% of respondents disagreed which might be caused by fear of information breach or manipulation.

*Sharing and Support Group*

Nowadays, people are connecting to many social networking sites. They share and discuss about many issues concerning with their own interest. In the study, we interested to figure out the effect of social media in healthcare settings. First, we asked to the respondents of their agreement on the ability of customers to share their experiences in the social networks in regard to the service they receive from healthcare providers, the study shows that 24% (strongly agree) and 48 % (agree) of respondents will share their experiences in dealing with healthcare providers the service they receive. They may think that sharing is a good way to improve health services. There are 18% (disagree) and 10% (strongly disagree), they belong to those respondents who view that social networks has nothing to do with healthcare services. This most likely caused by close management are practiced all health organizations for the time being.

Further, we asked that sharing between patients with similar condition in social networks can be beneficial and supporting each other. The survey reveals that most of the respondents (20% agree and 60% strongly agree) that they wanted to get in touch with other patients who have the similar condition with them through social networks. People with similar problems can easily share their experience and knowledge online, which may leads to support each other, at least morally, in facing their problems. Finally, we pointed out about supporting group through social networks will make them more resilient and confident to take a necessary act and decision regarding their health. It shows that 93% of respondents confirmed, while only 7% was disagreeing that supporting groups can be achieved via social networks.

*Extended services*

Customers' centric paradigm views those customers (or in this case patients) are partners in the healthcare process. However, many people still worry about the information quality and reliability provided by e-health services. This issue can be overcome by assigning an online health educator to ensure information flow within electronic channels is reliable. This section discusses the online health educator and its role. Firstly, we asked their agreement if healthcare organizations can provide e-health service guided by online health educators. Almost all participants have the same opinion that they agree (27% strongly agree and 56% agree) to utilize the service if it is offered.

Secondly, the question was an online health educator should respond effectively to any online queries without any delay. The result shows 32 % strongly agree and 60 % agree. It indicates that almost all participants expect that online health educators should support the existence of e-health and most importantly quick responses are required to make customers satisfy with the e-health services. While, 8% of respondents disagreed because of no clear rewards and compensation that they will receive in order to provide extra service to customers. In respond to the previous question, we are interested to find out the preference of customers whether they are willing to pay if an extra online service provided. Interesting facts show that less than half of the respondents agree to pay for the online health services (39%), while the majority (61%) prefer not to pay the service. Further, we asked about their agreement regarding the idea that physical health is also affected by emotional, psychological, and spiritual condition of the individual. Surprisingly, the result revealed that 65 % strongly agree with the statement, 35 % agree, and none of them disagree. This result will trigger any further research in this topic.

Finally, the last two questions asked were health literacy and customer satisfaction. Firstly, e-health activities such as access online medical record, online consultation and online discussion with other patients can improve health literacy. It shows 22% of respondents strongly agree, 42% agree, 30% disagree, and 6% strongly disagree. It is believed that embracing Web 2.0 in e-health service can improve their health literacy. Secondly, customers (patients) will be more satisfied with the service if a healthcare organization provides an online service where the patient is able to access their online medical record, online consultation, and online discussion with other patients. The result shows majority of respondents confirm (30% strongly agree and 57% agree) that those facilities will make them more satisfy if it is being offered. It signifies that e-health service with Web 2.0 tools is believed to improve their health literacy and customer satisfaction. The finding of this section will be used to recommend the implementation of Web 2.0 in e-health environment (Anshari, 2020).

## Discussion

This survey could be the first time in Indonesia focusing on Web 2.0 in e-health systems. Many e-health topics are covered, including online service, health information accessibility, and issues like social support and social networking. Data gathered from the survey will be used to formulate recommendations for the future direction of e-health and develop prototype of the systems. The 19 items in the survey instrument ware divided into three sections. The first section contains demographic traits of respondents to find out their gender composition, employment, and age segmentation, level of education, ICT literacy, and Internet literacy. Nineteen items in the second part are related to features of e-health services and its accessibility. In this section, we discuss findings and how they relate to the theories of e-health as discussed in the previous section.

E-health with Web 2.0 is not going to replace any existing services in healthcare. It will complement and extend the existing services with features that can improve quality of service. For example, it offers healthcare services more comprehensive and reliable and those services can be accessed online anytime-anywhere conveniently. The survey confirmed that the expectation in healthcare are high, which create challenges for any healthcare organization.

From the survey, we identify critical success factors that provide direction where e-health initiative should be heading. Managing those critical success factors are essential for a successful Web 2.0 initiative in the future e-health implementation. This research indicates that demographic pattern such as employment, age composition, education levels, ICT literacy, and Internet literacy that make up typical e-health initiatives are important parameters. It is interesting to see how demographic conditions affect the preference of respondents towards features of Web 2.0 in e-health services; this provides insight into their relative impact on e-health systems.

Respondents of the survey are fairly distributed, they were from both patients and healthcare staffs. Both types of respondents support the accessibility of medical records, empowerment, and social networks in e-health services. Level of education will affect the acceptability of systems and competency of people to utilize it. In this study two third of the respondents hold graduate diplomas, hence their education level are relatively high and this help them to easily grasp of e-health initiatives, including incorporating Web 2.0 on those initiatives.

The majority of the respondents is exposed to the Internet and they tend to use the Internet every day. Thus, e-health initiatives, including Web 2.0 services have the potential to be adopted as long as it addresses the need of people. This further supported that ICT literacy in urban are is relatively high and ICT literacy is an enabler factor of e-health initiatives. In this survey, we found that all respondents use computers on a daily basis, only a small percentage of the respondents use a computer a few times a week. This indicates that most of the population is computer literate and finding information or interacting using computers is not a problem.

In summary, the majority of respondents are educated and they are proficient in ICT as well. This is can be main reason as to why they support the implementation of an e-health system incorporating Web 2.0 features. The next section discusses our findings based on the survey in three categories: empowerment, social networks, and online health educator.

**Empowerment**

Empowerment is a characteristic of groups or individuals that energizes them with the knowledge and confidence to act in their own behalf in a manner that best meets their goals. A paradigm shift has been taking place in healthcare industry. The survey amplifies the trend of e-participation that they expect to secure and raise availability, accessibility and quality of health services. Empowerment is a result of both interactive and personal processes, where the emergence of 'power' (or potential) is facilitated by caring relationships (McWilliam et al., 1997). Web 2.0 is the enabler of e-empowerment and e-participation through electronic media (the Internet) as it provides convenient way to participate in decision making as well as controlling the flow of information.

Two questions directly asked about customers' empowerment in e-health services. First, we asked about the possibility of customers to access their medical records online, and secondly we asked to the respondents about features that enable customers (patients) to record and monitor their health habits online. It is important to note that though the respondents comprised of medical staff and non-medical staff (customers) evenly distributed, in an average 72 % of respondents supported the empowerment in e-health services. It indicates that majority expects to see the online services, which empower them to actively participate in the process of care.

However, those who disagree are mainly concern with the medium of medical records accessed. In principle, the contents of the medical record belong to patients; while the medical record file (physically) is the property of the hospital or health institution. Act No. 10 of Ministry of Health 749a states that any medical record file is owned health care organization, which should be kept at least for a period of 5 years from the date of the last patient treatment.

Because the patient owns the contents of medical records, a doctor or any medical staff can say about the content of the patient's medical record, except in certain circumstances that force doctors to act otherwise. Conversely, because a medical record file is owned by an organization, then a patient cannot access the file if the healthcare organization refuses the access

Customers' (in this case patients') empowerment means to support and encourage them to have more responsibility for their own health and to take a greater role in decisions about their own health care. The survey confirmed that patient preferred to have more ability and control of health information that concern with their health status. Making appointment online, viewing their medical records, and consulting online with medical staffs are few examples of patients' empowerment. The authorization to access health information will benefit both patient and healthcare organization. From a patient perspective, their health literacy is expected to improve by the time they have better knowledge of their own health status, and for the healthcare organization is expected to be a long lasting relationship since they always in need to access the service. Additionally, empowerment is also believed to give customers flexible time when and where they want to look upon their health related activities. Finally, the authors suggest that the features of empowerment in e-health system should exist in three dimensions; patient-patient interactions through social networks, self-patient empowerment in accessing their records in e-health systems, and patient-provider interaction through online health educators.

**Social Networks**

Some questions asked about social networks in healthcare environments. First, whether they want to share their experiences in the social networks regarding the service they receive from healthcare providers. Almost all respondents agreed that they would use the facilities of sharing with others if healthcare provider offers it. Secondly, the majority of respondents confirmed that sharing between patients with similar condition could be beneficial for them as they can share experiences about their conditions that might be helpful for others. People with similar problems can easily share their experience and knowledge online, which leads to supporting each other at least morally in facing their problems. Furthermore, support groups such as social networks in healthcare in Indonesia are not common. Respondents stated their endorsement on the possibility of support groups. A patient with who might have an illness that is socially shunned could find refuge in a social group with other patients who are conditioned as they are. In addition,

There must be a proper channel for customers to express their feeling regarding the service they receive from a healthcare provider. A healthcare provider through a social network site (preferably internal site) can provide this channel. Any concern or issue raised can be monitored and handled wisely. This can be a good feedback for the provider to improve its service to its patients.

The use of social networks is common in Indonesia since Indonesia is the 4th-largest user of the Twitter nation worldwide. However, adopting social networks into healthcare must be carefully designed to be able to get benefits for all parties not only to patients but also to healthcare providers, and the community as a whole. Based on the results of the survey, we propose that the provision of social networks or similar interaction channel where customers (patients) can express their satisfaction or dissatisfaction through the media that they maintain within their internal system. A healthcare provider can incorporate features of social networks into their e-health system whereby users who have authorization can take advantage of these facilities. This strategy can prevent undesirable situations that may jeopardize the image of the organization. With this in mind, the healthcare provider can accommodate Web 2.0 into their organizational strategy and practices. The survey results confirm the idea of Social CRM or

CRM 2.0 by Greenberg (2009) which concern that technology platform, business rules, processes, and social characteristics, designed to engage the customer in a collaborative conversation in order to provide mutually beneficial value in a trusted and transparent business environment. It's the organizational response to the customer's ownership of the conversation.

Catherine and Barbara (2008) mentioned that social networks and social supports closely affect to health outcomes. Therefore, we asked respondents whether they need facilities of social networks and social supports online. The majority of respondents agree that social networks and social support online as an integrated part of the e-health systems. It should operate, manage, and maintain within healthcare's infrastructure. This is more targeted to internal patients/families within the healthcare to have a conversation between patients/family within the same interest or health problem/ illness. For example, a patient with diabetic would motivate to share his/her experiences, learning, and knowledge with other diabetic patients. When patient/family is able to generates the contents of the Web, it can promote useful learning centre for others, not only promoting health among each others, but also it can be supporting group. Their experiences related to all issues such as; how the healthcare does a treatment, how much it will cost them, what insurance accepted by healthcare, how is the food and nutrition provided, etc. Furthermore, it also suggests that medical staff or doctors should be actively involved in an online forum discussion to provide professional advice. For healthcare management, conversation generates between patients in social network site can be a starting place to construct a business strategy. Competent healthcare staff within the managerial level to capture the conversation and listen what patients say about services provided should moderate the social networking site. The role of this task could campaign of healthy life for the society, which is not intended for commercial benefit for short term, but it is beneficial for the community.

**Online Health Educator**

An Online Health Educator (OHE) plays a pivotal role in the process of interaction between patient and a health provider. Patients should know how to use the service. E-health services which empower patients to have better knowledge and control over their health data, yet they need to communicate with their healthcare providers. In order to achieve the goal, the presence of the online health educator determines the success or failure of the implementation. It should be part of the implementation strategy to ensure that there are groups of staff to ensure that e-health service is managed in a professional way. Any health activities, which they can accomplish through an online service, they may not come physically to the hospital. For instant, a patient faced the daunting tasks of travelling, there is significant time spent in finding a parking space, time spent in the waiting room and then finally waiting for the prescription medicines. Though they do not come to the healthcare centre, yet they may need an assistant or guidance from online health educator when they face difficulties interacting with systems, which they need to discuss with online health educators for clarification and interpretation such as medical data, online consultation, or asking for any online service

In light of this, the results of survey confirm that the availability of OHE is important. Presence of OHE is the vital point in e-health instead of ICT as a tool. They are expected to have the skills to interpret medical data, able to guide patient go through technical systems, and know how to respond online queries properly. Interacting OHE with through Social RM in e-health promotion, education, and in emergencies would give advantages like early warning systems, especially in emergencies or pandemic. For instance, during the H1N1 flu outbreak, it is possible for health authorities to announce preventive measures over the e-health system to notify the public efficiently. In addition, this would also reduce the time-wasted in printing leaflets or posters to inform the public in such situations.

It is important to note that online communication is fast, yet cheap. Furthermore some people are more extrovert and more comfortable when communicating virtually. When patients are assigned to have a diet program, they do not have to travel all the way to the hospital because they can have access to the program using the online service provided and it saves time and money. Another benefit also for health professionals, they can monitor their patients' progress online, which is convenient for both parties. In addition, patients can easily share their health status after conducting their health program to their friends and relatives, which is good to gain moral support as well as to promote others to participate.

## Future Direction

The customer acceptance and demand for online healthcare services and technology among many segments of the general population has been high, there is ample evidence in the survey to support the prediction that the benefits of e-health through Web 2.0 technologies have the potential to make a profound impact in customer's expectation and all areas of healthcare services. Identifying critical success factors for Web 2.0 in e-health will provide direction of next generation of e-health systems The study points to the issue of demographic pattern such as employment, age composition, education levels, ICT literacy, and Internet literacy, as factors that make up typical e-health initiatives. It is important to see how those factors have affected the preference of e-health services. Secondly, the study revealed the significant components of empowerment, social networks, and online health educator to ensure the proposed prototype will be successfully implemented in accordance with the customers' expectation as derived from surveys. The prototype of a Web 2.0 application in a healthcare setting should adopt a concept of personalization of e-health service in which the Web is the platform that patient as partners in the healthcare process. They engage and involve in terms of generating content and interactions, as the patient is able to personalize his own interface in his profile page Web 2.0.

## Conclusion

Implementation of e-health will enable patient-friendly healthcare services and at the same time improve health provider's performance. At the national level, there is also a conscious move or effort by the government to understand the needs of the public and continuously upgrade service standards. In addition, the availability of Web 2.0 and the Internet infrastructure as well as the techno-savvy of the population will assist and accelerate the direction of e-health in the country. The respondents have responded with great support for the features and capability of e-health listed from the questionnaires. Unfortunately, there is a great a gap between available health services and services expected by the public, who seem to have advanced ICT literacy. From the survey, we can conclude that most respondents agree and are keen in having e-health services whilst requiring a high degree of confidentiality of their medical information. The majority of respondents support the list of abilities provided by an e-health system such as convenience in service, time saving, health promotion, accessibility, aware of medical result better, supporting each other, etc. Web 2.0 enables patients to communicate with the medical healthcare professional from conveniently, accessible anywhere-anytime. In summary, Web 2.0 in e-health service can provide wide range of online services that may empower patients to take a greater role of their own health.